\documentclass[fleqn,usenatbib]{mnras}

\usepackage{newtxtext,newtxmath}

\usepackage[T1]{fontenc}

\usepackage{graphicx}	
\usepackage{amsmath}	

\graphicspath{{./}{Figures/}}

\title[Atmosphere Loss by Aerial Bursts]{Atmosphere Loss by Aerial Bursts}

\author[Trierweiler \& Schlichting]{
Isabella L. Trierweiler$^{1}$\thanks{E-mail: isabella.trierweiler@astro.ucla.edu}
Hilke E. Schlichting,$^{2}$
\\
$^{1}$Department of Physics and Astronomy, University of California, Los Angeles, CA 90095, USA\\
$^{2}$Department of Earth, Planetary, and Space Sciences, University of California, Los Angeles, CA 90095, USA}

\date{Accepted XXX. Received YYY; in original form ZZZ}

\pubyear{2021}

\begin{document}
\label{firstpage}
\pagerange{\pageref{firstpage}--\pageref{lastpage}}
\maketitle

\begin{abstract}
We present a simple analytic description of atmospheric mass loss by aerial bursts and demonstrate that mass loss from aerial bursts becomes significant when the maximum impactor size that leads to an aerial burst rather than a ground explosion, $r_o$, is larger than the minimum impactor size needed to achieve atmospheric loss, $r_{min}$. For vertical trajectories, which give the most stringent limit, this condition is approximately satisfied when $\rho_o/\rho_i \gtrsim 0.4 v_e/v_\infty$, which implies atmospheric densities need to be comparable to impactor densities for impactor velocities that are a few times the escape velocity of the planet. The range of impactor radii resulting in aerial burst-induced mass loss, $r_o-r_{min}$, increases with the ratio of the atmosphere to the impactor density and  with the trajectory angle of the impactor. The range of impactor radii that result in aerial burst-induced mass loss and the atmospheric mass lost is larger in adiabatic atmospheres than isothermal atmospheres of equivalent total mass, scale height, and atmospheric surface density. Our results imply that aerial bursts are not expected to significantly contribute to the atmospheric mass-loss history of Earth, but are expected to play an important role for planets and exoplanets similar to Neptune with significant atmospheres. For Neptune-like atmospheres, the atmospheric mass ejected per impactor mass by aerial bursts is comparable to that lost by ground explosions, which implies that, for impactors following a Dohnanyi size distribution, overall loss by aerial busts is expected to exceed that by ground explosions by a factor of $(r_{ground}/r_{aerial})^{0.5}$. 
\end{abstract}




\section{Introduction}

The initial atmosphere of the Earth and other terrestrial planets was likely established by an interplay of diverse loss and delivery processes. Relevant processes that could deliver volatiles to a planet's atmosphere include, for example, degassing of volatiles during planetesimal impacts \citep[e.g.][]{abe1985,zahnle1988}, volcanic outgassing \citep[e.g.][]{craddock2009,elkins2012},
tectonic activity \citealt{zahnle2007}), and accretion of hydrogen-helium gas from the primordial gas disk \citep[e.g.][]{lammer2014,ginzburg2016}.

As volatiles build up in the atmosphere, depending on composition and other planet properties, atmospheric mass may be lost due to high energy radiation from a planet's host stars \citep[e.g.][]{kulikov2007,owen2013}, core-powered mass loss \citep{ginzburg2018,gupta2019}, or impacts \citep[e.g.][]{melosh1989,genda2003,zahnle1992,schlichting2015,wyatt2020}. 

Observed and inferred evidence for impacts of small planetesimals are plentiful within our own solar system. Impacts can explain variations in atmospheric mass between moons such as Titan, Ganymede and Callisto \citep{zahnle1992}, and may be responsible for the erosion of Mars' atmosphere \citep{melosh1989}. Evidence for Earth impacts can be seen in the presence of our Moon (\citealt{cameron1976}, \citealt{canup2001}), the cratering on the Moon and Earth, and the geochemical traces of the late veneer (\citealt{warren1999}, \citealt{walker2004}, \citealt{walker2009}). Aside from atmosphere loss, impacts are important to Earth's history as they are theorized to have delivered volatiles to the surface (\citealt{owen1995}, \citealt{dauphas2003}). They are one possible source of water for our oceans and for organic molecules \citep{chyba1990}, though the types of impactors (asteroid, comet, planetary embryo, etc.) that could have delivered this material are still under debate.

When considering atmospheric mass-loss by impacts, it is instructive to distinguish between giant impacts and planetesimal impacts. A giant impact is a planetary scale collision that can lead to global atmospheric mass-loss, whereas a planetesimal impact can only eject mass locally above the tangent plane.
Although giant impacts can eject significantly more mass in a single event than a planetesimal impact, planetesimal impacts are more efficient than giant impacts, since they produce more atmospheric loss per impactor mass \citep{schlichting2015}. 

Depending on impactor sizes, impact velocities and impact angles, volatiles may be added to or removed from growing planetary embryos by planetesimal impacts \citep[e.g.][]{shuvalov2009,schlichting2015,wyatt2020,sinclair2020}. Past works that have focused on evaluating the collective impact on atmospheric erosion by a population of planetesimal impactors of varying sizes have concentrated on ground explosions \citep{schlichting2015,wyatt2020}, i.e. impactors large enough that they hit the ground. In this paper we focus on planetesimal impactors that are significantly decelerated in the atmosphere \citep{shuvalov2014,shuvalov2016} and evaluate their contribution to atmospheric loss of terrestrial planets and Neptune-like planets and exoplanets.

To study the role of small impactors in atmospheric loss and evolution, we build on previous work by \cite{schlichting2015}, who derived a simple analytic atmospheric mass-loss model for planetesimal impacts, which successfully captured previous numerical results by \citet{shuvalov2009}. We expand on the work by \cite{schlichting2015} first by accounting for impactors which do not reach the ground and instead explode as aerial bursts, and second by including both isothermal and adiabatic atmospheric profiles in our model. 


By reasoning that aerial bursts deposit most of their energy at a location in the atmosphere where they reach roughly half of their initial velocity, we calculate the atmospheric properties required for a planet to lose atmospheric mass by aerial bursts. We demonstrate that aerial burst-induced mass loss is only significant for planets with atmospheric densities comparable to the impactor density and quantify the resulting atmospheric mass loss by applying our model to Neptune-like planets and exoplanets.

This paper is structured as follows: In section \ref{methods} we outline our approach for calculating atmosphere loss by considering a simple impactor explosion model. In section \ref{fireball}  we apply our model to aerial bursts and present results for Neptune-like atmosphere in section \ref{neptune}. We outline general planet parameters for which mass loss by aerial bursts is important in section \ref{constraints}. In section \ref{ground} we discuss the contributions of aerial bursts to the total atmospheric mass loss. Finally, we summarize our findings in section \ref{conclusions}.

\section{Methods}{\label{methods}}
\subsection{Atmospheric Profiles}
Before presenting our mass loss model, we first describe the two atmospheric profiles assumed throughout this work, and the mass loss limits associated with each profile. We consider isothermal and adiabatic atmospheres with the following density profiles:
\begin{equation}
\begin{split}
    \rho(z) &= \rho_o e^{-z/h} \mbox{\quad (Isothermal)}\\
    \rho(z) &= \rho_o \left( 1 - \frac{\gamma-1}{\gamma} \frac{z}{h}\right)^{1/(\gamma-1)} \mbox{\quad (Adiabatic)}
\end{split}
\end{equation}
Here $h= \frac{k_B T}{\mu m_p g}$ is the scale height, where $k_B$ is the Boltzmann constant, $T$ is the temperature, $\mu$ is the mean molecular weight, $m_p$ is the proton mass, and $g = \frac{GM_p}{R^2}$ is the gravitational acceleration for a planet of mass $M_p$ and radius R. The atmospheric density $\rho$ is a function of $z$, the height in the atmosphere above the ground. $\rho_o$ is the density of the atmosphere at the planet's surface and $\gamma$ is the adiabatic index. Throughout this work we assume a diatomic atmosphere with $\gamma = 1.4$ so that the adiabatic profile is $\rho(z) = \rho_o \left( 1 - \frac{2}{7} \frac{z}{h}\right)^{5/2}$. The atmosphere extends from $z=0$ to $z= \infty$ for the isothermal profile and to $z = 7h/2$ for the diatomic, adiabatic profile.

\begin{figure}
\centering
    \includegraphics[width=0.49\textwidth]{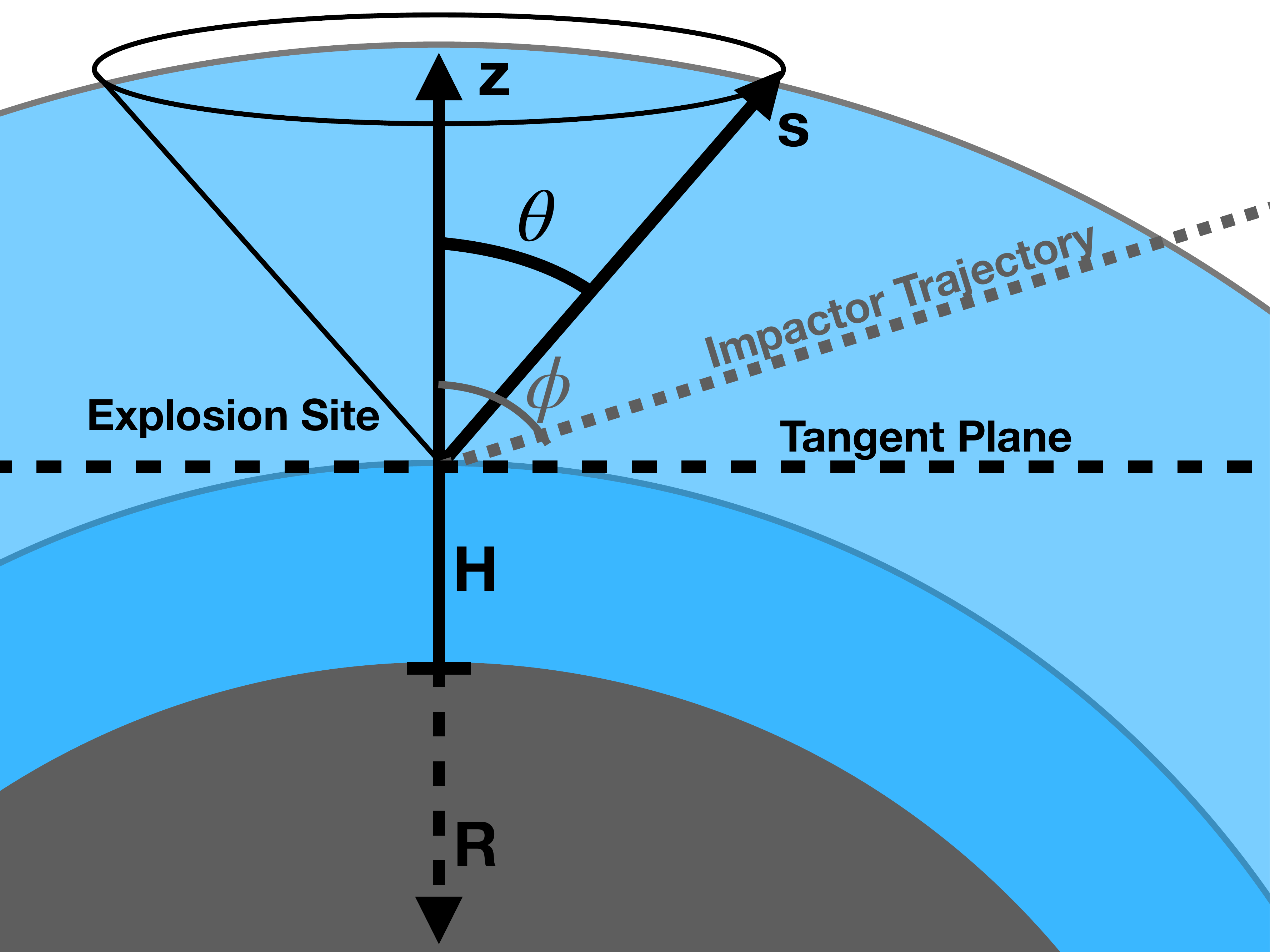}
   \caption{\textbf{Diagram of an aerial burst.} The planet has a radius $R$ and an atmosphere with a scale height $h$. The planetesimal enters the atmosphere along the grey dashed line with an angle $\phi$ relative to the normal above the explosion site. It explodes as an aerial burst at height $H$ above the ground. From the explosion site, we define $z$ to be the height above the explosion ($z=0$ at the height $H$), and $s$ to be a column of atmosphere at an angle of $\theta$ from $z$. For an aerial burst occurring at height $H$, the maximum mass of atmosphere which can be ejected is roughly the cap of atmosphere out to the tangent plane ($\theta = \pi/2$). }\label{coordinates_1}
\end{figure}

Using a coordinate system centered on the impact site (see figure \ref{coordinates_1}), and assuming $h\ll R$, we can write our coordinates as $s = \sqrt{R^2 \cos^2\theta + 2zR} - R\cos\theta$ or $z = (s^2 + 2Rs\cos\theta)/2R$. We note that the assumption of $h\ll R$ means that the density of the atmosphere will decrease quickly with height, so that the density profile will be close to zero for heights of $z \gg R$. All aerial bursts will therefore occur at heights $H\ll R$. An additional benefit of the assumption $h\ll R$ is that the height of the aerial burst $H$ only factors in the density profile, and not in $z$ or $s$, simplifying many of the following equations. 

Assuming $h\ll R$, the total mass in the atmosphere for both the isothermal and adiabatic profiles is well approximated by \begin{equation}
    M_{tot} \simeq 4 \pi R^2 \int_0^{z(\rho\rightarrow0)} \rho(z) dz = 4 \pi \rho_o h R^2.
\end{equation}

The maximum mass of atmosphere that can be ejected by an aerial burst at height $H$ is roughly the cap of atmosphere above the tangent plane. We call this mass $M_{cap}(H)$, and calculate it as

\begin{equation}
    M_{cap} (H)= 2 \pi \int_0^{\pi/2} \int_0^{z(\rho\rightarrow0)} \rho(z+H) s^2 \sin(\theta) ds d\theta.
\end{equation}
For an explosion on the ground ($H= 0$), the cap mass for each profile is $M_{cap, Iso} = 2 \pi \rho_o h^2 R$ and $M_{cap, Adi} = \frac{14}{9} \pi \rho_o h^2 R$.

\subsection{Mass Loss Model}
We model each aerial burst as an explosion at the height in the atmosphere where drag forces significantly decelerate the body. Specifically, we choose the height where drag forces slow the impactor to half of its initial velocity, such that the impactor's momentum is halved and its kinetic energy is $1/4$ of the initial kinetic energy. The assumption that the half velocity height is the explosion height is a reasonable approximation because the atmospheric density increases exponentially (or as a steep power-law in the adiabatic case) as the body approaches the surface such that it experiences most of its deceleration around this half-velocity height.

The direction of the impact trajectory is parameterized by the angle $\phi$, which is measured from the normal above the explosion site (see figure \ref{coordinates_1}), such that $\phi=0$ represents a vertically falling impactor. The impactors' initial velocity is $v_\infty$. We account for impactors approaching the planet at any angle, so that the half-velocity height is a function of both the radius of the impactor and the trajectory angle $\phi$. We outline this half velocity calculation in section \ref{halfv}. Smaller impactors are decelerated to half their initial velocity higher in the atmosphere than larger impactors. Additionally, impactors travelling on nearly horizontal trajectories will pass through more atmosphere than those following vertical trajectories, and so the horizontal impactors will have relatively higher half-velocity heights. Bodies which decelerate to half their initial velocity before hitting ground are referred to as aerial bursts, while those that reach the ground with more than half of their initial velocity are referred to as ground explosions.

At the calculated explosion height, we model the effect of each impactor as a spherically symmetric point explosion. As the impactor travels through the atmosphere, it creates a shock which heats the gas ahead of it (\citealt{zeldovich1967}, \citealt{vickery1990}). When the impactor decelerates, either by hitting the planet's surface or through drag forces, the heated gas expands as a vapor plume. The plume expands in the atmosphere, transferring momentum to the surrounding gas. Any gas in the plume that is accelerated to the escape velocity of the planet is then ejected. This expanding plume can be well modeled by an explosion placed at the point where the impactor deposits most of its energy in the atmosphere (\citealt{vickery1990}). The explosion is modeled as an isotropic sphere of the same mass as the impactor, expanding at the same velocity. The energy of this expanding mass is then transferred to the surrounding atmosphere. 

In each explosion, we distribute the energy of the impactor over the full sphere around the half-velocity point. In this model, the atmospheric mass that will be ejected is the cone of atmosphere which can be accelerated by the impactor's explosion to the planet's escape velocity. We define the maximum angle along which atmosphere can be ejected to be $\theta_m$. Assuming energy conservation, this angle is where the atmospheric mass per solid angle is equal to impactor mass ($M_i = \frac{4 \pi}{3} r^3 \rho_i$) distributed over a sphere and scaled by the final and escape velocities:
\begin{equation}\label{mami}
    \int_0^{s(\rho\rightarrow0)} \rho(z+H) s^2(\theta_m) ds = \frac{M_i}{4 \pi} \frac{v_f^2}{v_e^2}.
\end{equation} 
Here $v_f$ is the velocity of the impactor when it reaches the explosion site, $v_e$ is the escape velocity of the planet, and $s(\theta_m) = \sqrt{R^2 \cos^2\theta_m + 2zR} - R\cos\theta_m$. 

We generally assume that the impactors are rocky bodies approaching the planet at $2 v_e$, the velocity we would expect for a body such as an asteroid. We also consider comet-like bodies, which have lower densities and expected initial velocities around $5 v_e$.

The total mass lost is then found by integrating over the cone of material out to $\theta_m$
\begin{equation} \label{macc}
    M_{eject} = 2 \pi \int_0^{\theta_{m}} \int_0^{s(\rho\rightarrow0)} \rho(z+H) s^2 \sin\theta ds d\theta.
\end{equation}

\subsection{Half Velocity Heights}\label{halfv}
 In the point explosion model for impactors, we assume the explosion occurs at the height where the impactor is slowed to half its initial velocity by drag forces. To calculate this half velocity height, we calculate the velocity of the body as a function of the height of the impactor above the ground. 

Bodies will always approach a planet with at least the escape velocity, and asteroid and comet-like bodies are expected to be travelling at even higher speeds. Because we are considering atmospheres for which $h\ll R$, any significant deceleration the impactor experiences due to gas drag will occur over a relativity short distance compared to R and within heights significantly smaller than R above the ground. As such, in calculating the impactor's velocity we account for the drag force but neglect any additional gravitational acceleration the impactor experiences as it is always either too weak or acting on too short of a timescale to significantly change the body's velocity before the body reaches $1/2 v_\infty \geq v_e$. We can verify that this approximation is justified by taking the ratio of the drag force and gravitational force, $(R/r)(\rho_o/\rho_i)(v_\infty/v_e)^2$, and confirming that it is indeed greater than $\sim 1$, usually much greater, for the atmospheric densities and aerial burst impactor sizes investigated here.

Drag forces will flatten the bodies as they travel through the atmosphere, so we add a parameter $\alpha$ to the area of the body $(\rm Area = \alpha \pi r^2)$. While the body should flatten more over the course of its trajectory through the atmosphere, we will set $\alpha$ to be a constant for now. Based on numerical simulations (\citealt{shuvalov2014}), assuming an increase in radius of about 3 is reasonable, resulting in $\alpha =9$.
We note that while this flattening is due to fragmentation of the body, the impactor is still expected to continue travelling as a single entity. Smaller fragmenting pieces will fill in the gaps between larger fragments, allowing us to model the impactor as a falling pancake \citep{hills1993}. With the $\alpha$ parameter for flattening, the equation for the drag force acting on the impactor becomes

\begin{equation}\label{Fd}
    F_d = \frac{1}{2} C \rho(z) \alpha \pi r^2 v^2,
\end{equation}

where $\alpha \pi r^2$ is the impactor's effective cross section, v is the velocity, r is the impactor radius, and $C$ is the drag coefficient. For a sphere, $C =0.5$, however the flattened shape of the impactor will change the drag coefficient. Coefficients for non-spherical, rough bodies at high speeds can reach 1 or higher, so we will assume $C =1$.

The equation of motion for the impactor is therefore 
\begin{equation}\label{vvinf1}
    \frac{vdv}{ds} = -\frac{1}{2} \alpha C \rho(z) \pi r^2 v^2 / M_i. 
\end{equation}

We note that the impactor's velocity will depend on the angle $\phi$ of the impactor's trajectory through the atmosphere, measured relative to the vertical axis $z$ (see figure \ref{coordinates_1}). For a given $\phi$, we can solve equation \ref{vvinf1} for the velocity of the body at a height $H$ above the ground, finding

\begin{equation}\label{vvinf}
    \ln \left[\frac{v(H, \phi)}{v_\infty}\right] = - \frac{3}{8} \frac{\alpha C}{r \rho_i}\int_{s(\rho\rightarrow0)}^{0} \rho(z+H) ds, 
\end{equation}
where $\rho_i$ is the density of the impactor, $s = \sqrt{R^2 \cos^2\phi + 2zR} - R\cos\phi$, and $\phi$ is the angle from the vertical along which the impactor moves through the atmosphere. 

Finally, defining the height $H$ to be where the velocity of the body slows to half of its initial velocity ($v(H, \phi)=\frac{1}{2} v_\infty$), we find 
\begin{equation}\label{halfvint}
\begin{split}
    \ln(1/2) &= - \frac{3}{8} \frac{\alpha C}{r \rho_i}\int_{s(\rho\rightarrow0)}^{0} \rho(z+H) ds\\
    &\mbox{or equivalently, }\\
    \ln(1/2) &= - \frac{3}{8} \frac{\alpha C}{r \rho_i}\int_{z(\rho\rightarrow0)}^{0} \frac{ \rho(z+H) R}{\sqrt{R^2 \cos^2\phi + 2zR}}dz, 
\end{split}
\end{equation}
where $H$ is the half velocity height. 


For trajectory angles $\phi \ll \pi/2$ (i.e., vertical trajectories close to the $z$ axis), we can approximate the half velocity height in equation \ref{halfvint} as a function of the trajectory angle $\phi$, and the radius of the body $r$:
\begin{equation}\label{halfvheight}
\begin{split}
    \frac{H}{h}_{Iso} (r, \phi) &= \ln\left(\frac{3}{8} \frac{\alpha C}{\ln2} \frac{h}{r} \frac{\rho_o}{\rho_i} \frac{1}{\cos\phi}\right)\\
    \frac{H}{h}_{Adi} (r, \phi) &= \frac{7}{2} \left[1 - \left(\frac{8}{3} \frac{\ln2}{\alpha C} \frac{r}{h} \frac{\rho_i}{\rho_o} \cos\phi\right)^{2/7} \right], 
\end{split}
\end{equation}
where $h$ is the scale height, the densities of the atmosphere and impactor are $\rho_o$ and $\rho_i$, respectively, $C$ is the drag coefficient, and the $\alpha$ is the flattening parameter.

\begin{figure}
    \centering
    \includegraphics[width=0.49\textwidth]{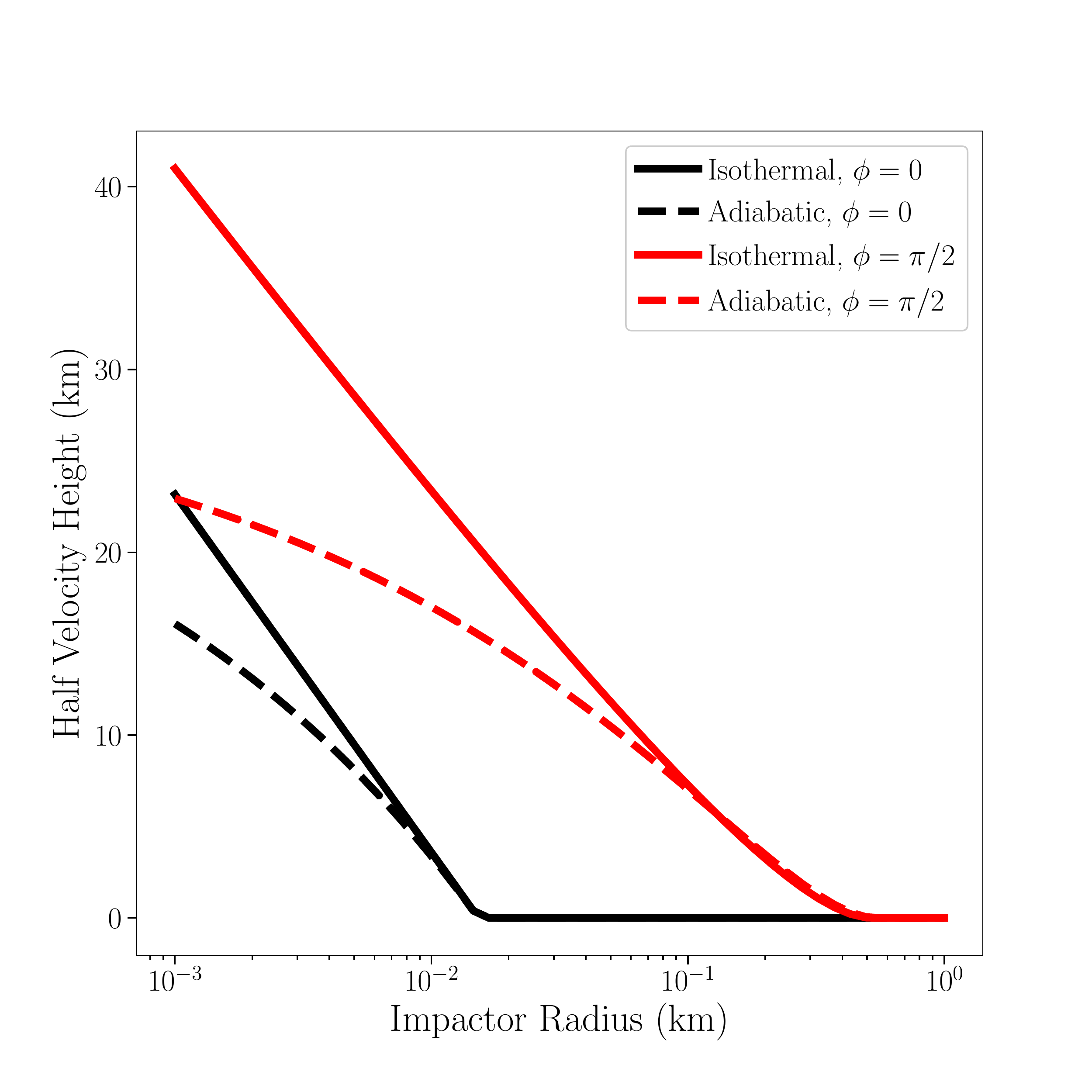}
    \caption{\textbf{Calculated half velocity heights as a function of impactor radius} (equation \ref{halfvheight}), for asteroids ($\rho_i = 3300$ kg/m$^3$, $v_\infty = 2v_e$) impacting an Earth-like atmosphere ($h=8$ km, $\rho_o= 1.2$ kg/m$^3$). The solid curves show the isothermal profile, while the dashed curves show the adiabatic profile. The black curve is calculated for a vertical impactor trajectory ($\phi = 0$), and the red curve is for a horizontal trajectory ($\phi = \pi/2$). } \label{halfheight}
\end{figure}

Full numerical solutions for the half velocity heights in isothermal and adiabatic profiles, illustrating two different $\phi$ angles, are found in figure \ref{halfheight}. These heights are calculated for an Earth-like atmosphere. We see that larger impactors have lower half velocity heights. Comparing the explosion heights for the vertical ($\phi=0$) and horizontal ($\phi = \pi/2$), and isothermal and adiabatic cases, we see that the heights for horizontal trajectories are higher by up to a factor of two for an Earth-like atmosphere. This difference is because impactors on horizontal trajectories must pass through more atmosphere than those on vertical trajectories. Impactors in an adiabatic atmosphere also have an upper limit on the half velocity height as the atmosphere extends only to $z=7h/2$.

\subsection{Comparing analytical and numerical half velocity heights}
\begin{figure}
    \centering
    \includegraphics[width=0.49\textwidth]{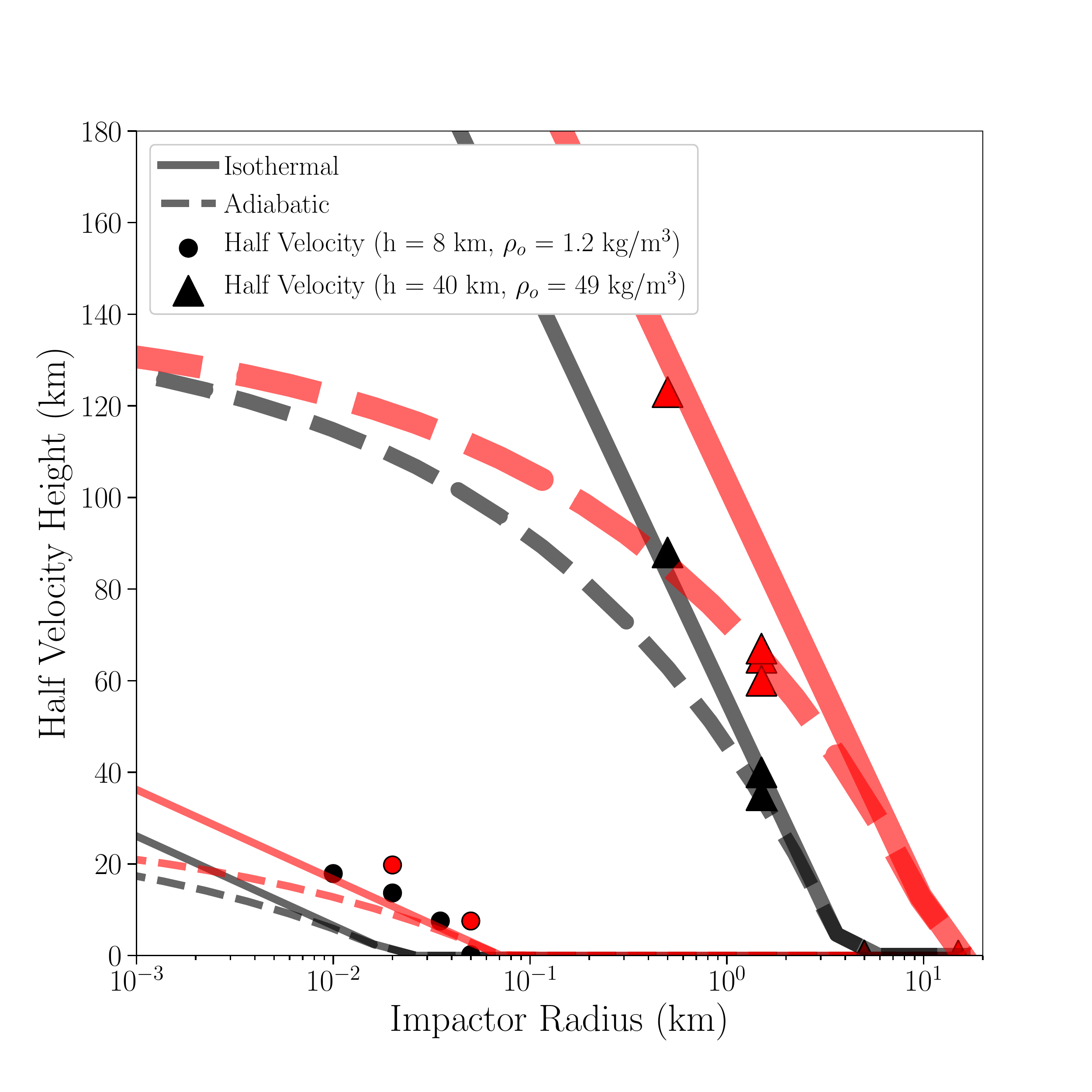}
    \caption{\textbf{Comparison of analytical half velocity heights to those calculated in hydrodynamics simulations}. The analytical heights are calculated using equation \ref{halfvheight} and the simulated heights are from \citealt{shuvalov2014} and \citealt{shuvalov2016}. The circular points are from \citealt{shuvalov2016} and correspond to numerical heights calculated for an Earth-like atmosphere ($h= 8$ km, $\rho_o = 1.2$ kg/m$^3$). The triangular points are from \citealt{shuvalov2014} for a primordial Earth atmosphere ($h= 40$ km, $\rho_o = 49$ kg/m$^3$). The solid curves are calculated using an isothermal profile while the dashed lines are calculated with an adiabatic profile. The thin curves are the analytical heights for the Earth atmosphere while the thick curves are for the primordial Earth atmosphere. The red curves and points represent impactors with comet properties ($\rho_i = 1000$ kg/m$^3$) while the black curves and points are asteroid-like impactors ($\rho_i = 3300$ km/m$^3$). We find that the numerical half velocity heights for both the Earth and primordial Earth atmospheres are in reasonable agreement with the analytical curves, for both comet and asteroid impactors. }\label{heights_data}
\end{figure}

We compare our derived half velocity heights with those obtained by \citet{shuvalov2014,shuvalov2016} using hydrodynamical simulations (figure \ref{heights_data}). These works investigated impactors entering two types of atmospheres - one analogous to Earth's current atmosphere ($h= 8$ km, $\rho_o = 1.2$ kg/m$^3$) and one representing a primordial atmosphere ($h= 40$ km, $\rho_o = 49$ kg/m$^3$). The two works included simulations for both asteroid and comet impactors, with a variety of assumed initial velocities. 

In figure \ref{heights_data}, we compare the half velocity heights calculated in the numerical simulations \citep{shuvalov2014,shuvalov2016} to our analytical half velocity heights. The circular points indicate numerical heights calculated for the $h = 8$ km atmosphere while the triangular points represent the numerically calculated heights for the denser, $h= 40$ km atmosphere. Red points represent simulations of comets ($\rho_i = 1000$ kg/m$^3$) and black points represent asteroids ($\rho_i = 3000$ kg/m$^3$). Due to the sensitivity of the numerical simulations to initial conditions, some impactor radii have multiple corresponding half velocity heights.

The solid and dashed curves in figure \ref{heights_data} are the analytical half velocity heights described by equation \ref{halfvheight}. The solid curves are calculated assuming an isothermal atmosphere, while the dashed curves are for adiabatic atmospheres. The thicker curves are heights calculated for the $h= 40$ km, $\rho_o = 49$ kg/m$^3$ atmosphere while the thin curves at lower heights are for the $h=8$ km, $\rho_o = 1.2$ kg/m$^3$ atmosphere. As with the numerical points, the red curves correspond to half velocity heights for comets while the black curves correspond to the half velocity heights calculated for impactors from the asteroid belt. 
We find that our analytical expression for the half velocity height is in reasonable agreement with numerical results for both type of atmospheres as well as for cometary and asteroidal impactors.

In our mass loss model described below, we will assume that the half velocity height is the location of the point explosion.


\section{Mass loss by aerial bursts}\label{fireball}
The aerial burst atmospheric mass loss regime includes any impactors which can eject atmosphere and which slow to half their initial velocity before reaching the planet's surface. As described in section \ref{methods}, aerial bursts are modeled as explosions at the height $H$ above the ground. At this height, the expanding mass is distributed over the whole sphere around the explosion point. 


To calculate the total mass loss due to aerial bursts, we first pick a height $H$ and trajectory angle $\phi$ between $0$ and $\pi/2$, and then calculate the impactor radius corresponding to this explosion height given the trajecotry angle (equation \ref{halfvheight}). Following the explosion model outlined in section \ref{methods}, we then calculate the maximum angle along which an impactor of the calculated radius can eject atmosphere ($\theta_m$ in equation \ref{mami}). Noting that the final velocity of the aerial burst is half of its initial velocity ($v_f = v_\infty/2$), we write equation \ref{mami} as: 

\begin{equation}
    \int_0^{s(\rho\rightarrow0)} \rho(z+H, z = s^2/2R + s \cos\theta_m) s^2 ds = \frac{r^3}{12} \rho_i \frac{v_\infty^2}{v_e^2}.
\end{equation}

Using this angle $\theta_m$ and equation \ref{macc}, we calculate the mass ejected. This yields mass loss as a function of impactor radius and trajectory angle at a given height. We then sum over the explosion heights in order to find the overall expected mass loss as a function of impactor radius and trajectory angle. 

\subsection{Characteristic Radii}\label{characteristic radii}
Two limiting impactor radii define the bounds of atmospheric mass loss by aerial bursts. $r_o(\phi)$, which describes the impactor radius corresponding to an explosion height of zero for a given trajectory angle, sets the upper limit for the radius of aerial burst projectiles. At radii larger than $r_o(\phi)$, impactors will result in ground explosions. $r_{min} (\phi)$, the minimum impactor radius that can eject mass from a given trajectory angle, sets the lower radius limit for planetesimals that will result in any mass loss. As such, aerial burst mass loss can occur only when $r_{min} < r < r_o$, where $r$ is the planetesimal radius. 

Starting with the equations relating the half velocity height, trajectory angle, and impactor radius (equation \ref{halfvheight}), we solve for the impactor radius where the explosion height is zero. Rewriting equation \ref{halfvheight} for the case where $r_o(\phi) = r(\phi)|_{H=0}$ gives

\begin{equation}\label{rground_int}
    \frac{r_o}{h}(\phi) = \frac{3}{8} \frac{\alpha C}{\ln2} \frac{1}{\rho_i} \int_{z(\rho\rightarrow0)}^{0} \frac{ \rho(z) R}{\sqrt{R^2 \cos^2\phi + 2zR}}dz.
\end{equation}

In the  limit where $\phi \ll \pi/2$, we find
\begin{equation}\label{rground}
    \frac{r_o}{h}_{Iso, Adi} (\phi) = \frac{3}{8} \frac{\alpha C}{\ln2} \frac{\rho_o}{\rho_i}\frac{1}{\cos\phi}.
\end{equation}
Note that the approximated form of $r_o(\phi)$ is equivalent for both the isothermal and adiabatic profiles.  

$r_{min}$ is calculated by considering the smallest planetesimal that will be able to eject mass from it's associated explosion height. At a given height, the minimum mass that may be ejected in an explosion is the column of atmosphere directly above the explosion site, so we first solve equation \ref{mami} for the impactor radius required to eject mass along the vertical $z$ axis, or equivalently along $\theta_m = 0$. With the condition that $s(\theta_m=0) =z$, we can rewrite equation \ref{mami} as

\begin{equation}\label{eq:rmin_integral}
\int_0^{z(\rho\rightarrow0)} \rho(z+H(r_{min}, \phi)) z^2 dz = \frac{\frac{4\pi}{3}\rho_i r_{min}^3}{4 \pi} \frac{v_f^2}{v_e^2},
\end{equation}
where $r_{min}$ is the radius required to eject the column of mass directly above an explosion site at height $H$, $\phi$ is the trajectory angle relative to the $z$ axis, $\rho_i$ is the impactor density, $v_f$ is the final velocity of the body at height $H$, and $v_e$ is the escape velocity of the planet. As defined in section \ref{methods}, the explosion height $H$ depends on both the radius of the planetesimal and the trajectory angle $\phi$. 

Substituting the expression of half velocity height for $H$, with a planetesimal radius of $r_{min}$ and a trajectory angle of $\phi$ ($H= H(r_{min}, \phi))$ in equation \ref{halfvheight}, and noting that $v_f = \frac{1}{2} v_\infty$ at the half velocity height, we calculate $r_{min}(\phi)$ for $\phi \ll \pi/2$ as

\begin{equation}\label{rmin}
\begin{split}
    \frac{r_{min}}{h}_{Iso} (\phi) &= \sqrt{64 \frac{\ln2}{\alpha C} \frac{v_e^2}{v_\infty^2} \cos\phi} \\
    \frac{r_{min}}{h}_{Adi} (\phi) &= 12^{7/10} \left(\frac{v_e}{v_\infty}\right)^{7/5} \left(\frac{\rho_i}{\rho_o}\right)^{2/5} \left(\frac{8}{3} \frac{\ln2}{\alpha C} \cos\phi\right)^{11/10}.
\end{split}
\end{equation}

We note that for a given total atmospheric mass, $r_{min}(\phi)$ in an adiabatic atmospheric profile will be somewhat lower than the $r_{min}(\phi)$ for the isothermal case. In other words, adiabatic aerial burst mass loss can occur at smaller impactor radii than isothermal aerial burst mass loss. 




\section{Application to sub-Neptunes}
\begin{figure}
    \centering
    \includegraphics[width=0.49\textwidth]{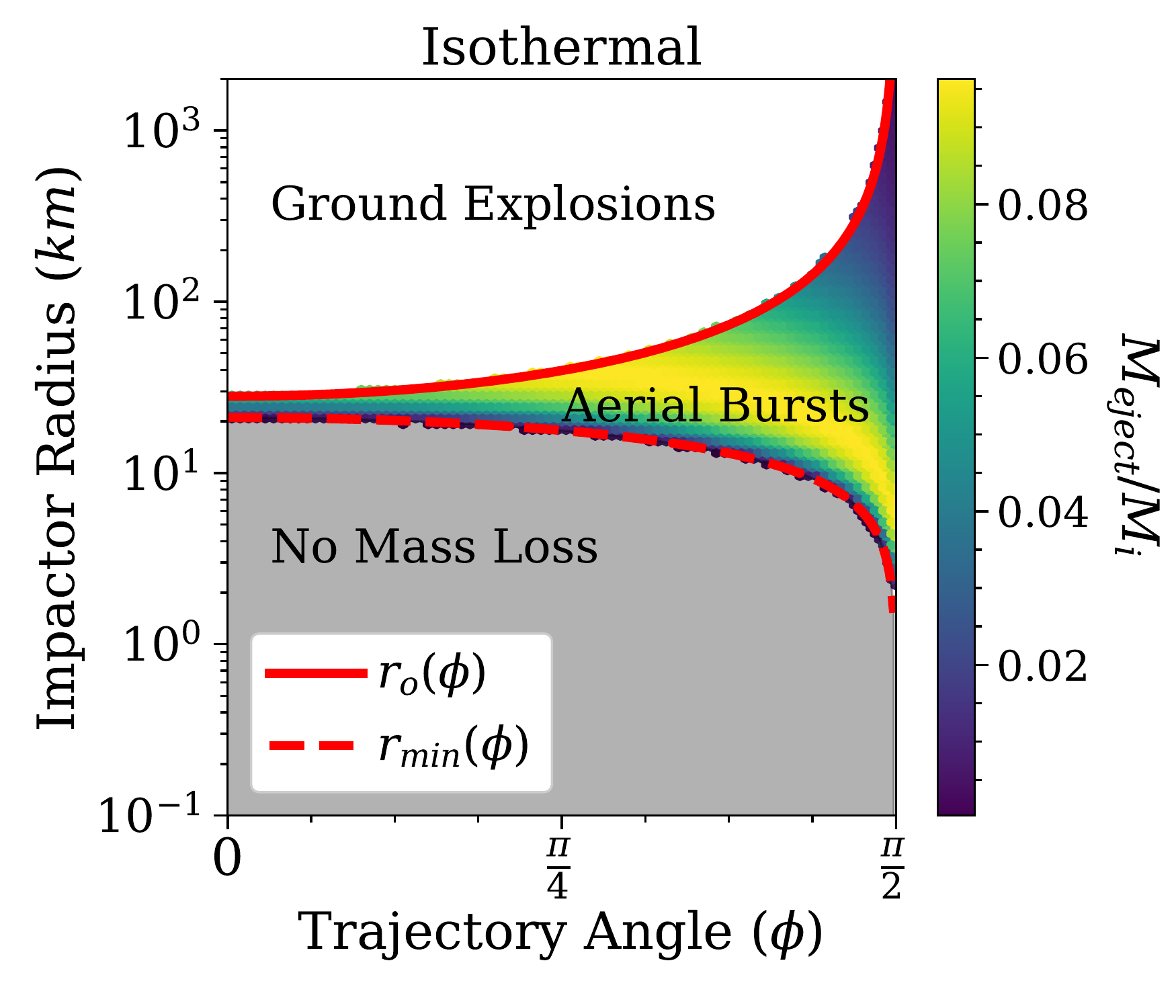}
    \includegraphics[width=0.49\textwidth]{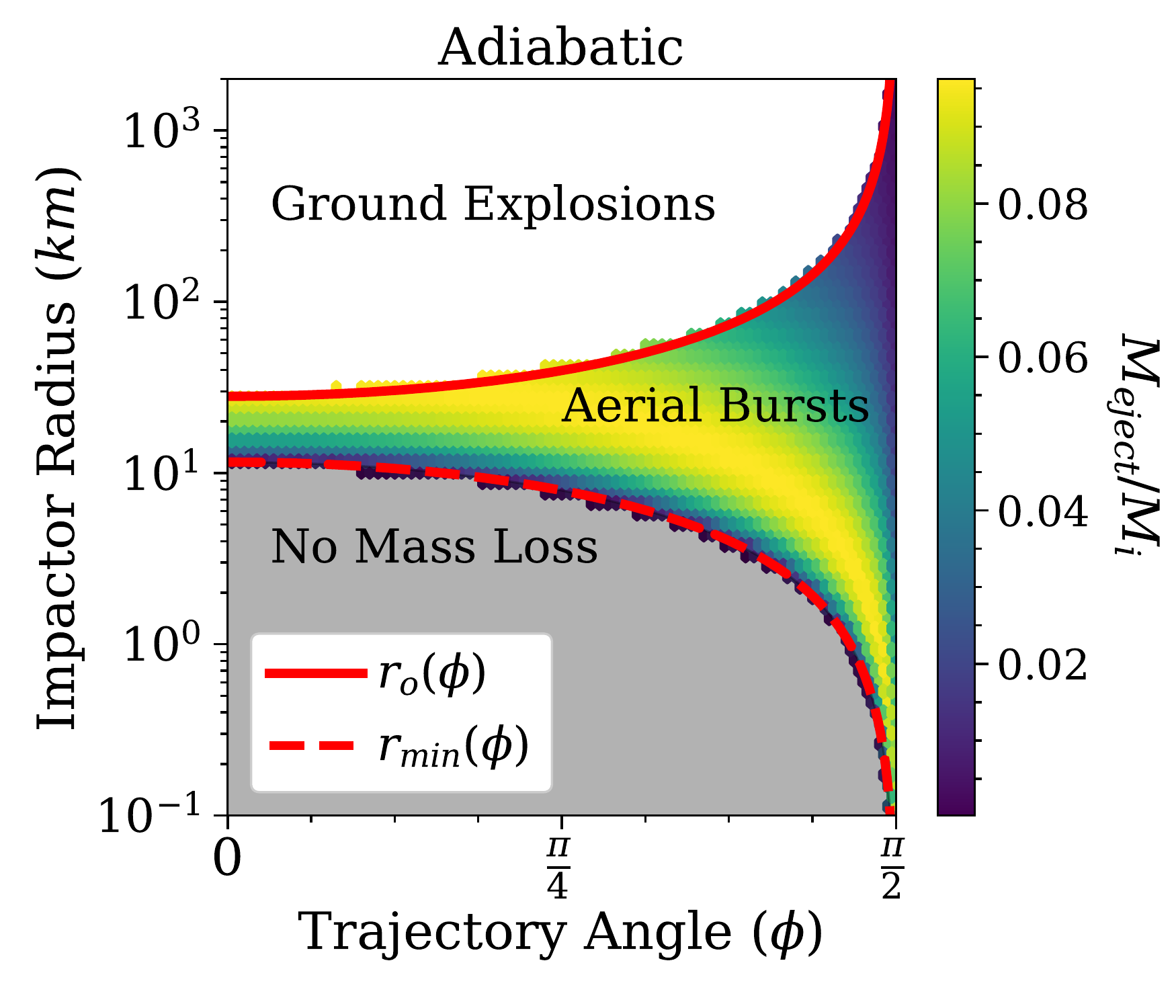}
   \caption{\textbf{Ejected atmospheric mass per impactor mass} as a function of the impactor radius and trajectory angle for asteroid-like impactors entering a sub-Neptune/Neptune like atmosphere. The trajectory angle $\phi$ is measured relative to the vertical $z$ axis (see figure \ref{coordinates_1}). The top figure shows mass loss assuming an isothermal atmosphere and the bottom figure shows mass loss for an adiabatic atmosphere. The solid curves show $r_o$ as a function of trajectory angle and the dashed curves show $r_{min}$. Aerial burst mass loss occurs between $r_{min}$ and $r_o$. Below $r_{min}$ (below the dashed line) no mass loss can occur. We assume an atmospheric density at the surface of $\rho_o\sim 1000$ kg/m$^3$ and an impactor density of $\rho_i = 3300$ kg/m$^3$. Mass loss by aerial bursts can occur at all trajectory angles, and occurs at relatively smaller impactor radii for the adiabatic case than the isothermal case.  }\label{neptune_fireballsonly}
\end{figure}

{\label{neptune}}
To illustrate the effects of atmospheric mass-loss by aerial bursts, we apply our model to a theoretical sub-Neptune atmosphere (figure \ref{neptune_fireballsonly}). We assume atmosphere properties similar to Neptune, with a scale height of 19 km, an atmospheric density of $\rho_o \sim 1000$ kg/m$^3$ at the base of the envelope, and a planetary core radius of $R=6000$ km. We consider asteroid-like impactors ($\rho_i = 3300$ kg/m$^3$) with an initial velocity of $v_\infty = 2v_e$. If the impactor density or initial velocity is greater than these values, mass loss would increase in magnitude at all impactor radii. 

Figure \ref{neptune_fireballsonly} shows the ejected atmosphere per unit impactor mass as a function of the projectile radius $r$ and the trajectory angle $\phi$ relative to the vertical $z$ axis. For comparison, we also show the approximated forms of $r_o(\phi)$ (solid curve) and $r_{min} (\phi)$ (dashed curve), as derived in section \ref{characteristic radii}. Aerial burst-induced mass loss occurs for the radii $r$ where $r_{min}<r<r_o$. At radii larger than $r_o$, impactors result in mass loss through ground explosions instead of aerial bursts (see section \ref{ground}).

A typical sub-Neptune/Neptune like planet has a large range of radii for which aerial burst-induced mass loss occurs, with most aerial bursts resulting in atmospheric mass loss equal to a few percent of the projectile mass. Because projectiles pass through relatively more atmosphere, and therefore slow to half velocity at higher heights, along horizontal ($\phi = \pi/2$) trajectories than vertical trajectories, there is a much larger range of planetesimal radii that can result in aerial burst-induced mass loss at large trajectory angles than at small angles. Due to the differences in atmospheric mass distribution in adiabatic and isothermal atmosphere profiles, aerial burst mass loss begins at relatively lower radii in the adiabatic model than isothermal model. 

\section{Constraints on the aerial burst Regime}\label{constraints}
Recall that at a given trajectory angle $\phi$, aerial burst-induced mass loss occurs for impactor radii greater than $r_{min} (\phi)$ and less than $r_o(\phi)$. In other words, projectiles must be massive enough to provide adequate energy for mass loss without being so large that they reach the planet surface without decelerating. As shown in section \ref{characteristic radii}, these radii depend on the atmospheric profile. Therefore, our definitions of $r_o$ and $r_{min}$ can be used to set limits on the properties of atmospheres that may lose mass through aerial bursts. Because aerial burst-induced mass loss occurs for planetesimal radii $r$ where $r_{min} < r < r_o$, for aerial burst mass loss to be possible in an atmosphere with a given atmospheric surface density $\rho_o$, $r_o(\phi)$ must be greater than $r_{min}(\phi)$ for at least some values of $\phi$. 

\begin{figure}
    \centering
    \includegraphics[width=0.49\textwidth]{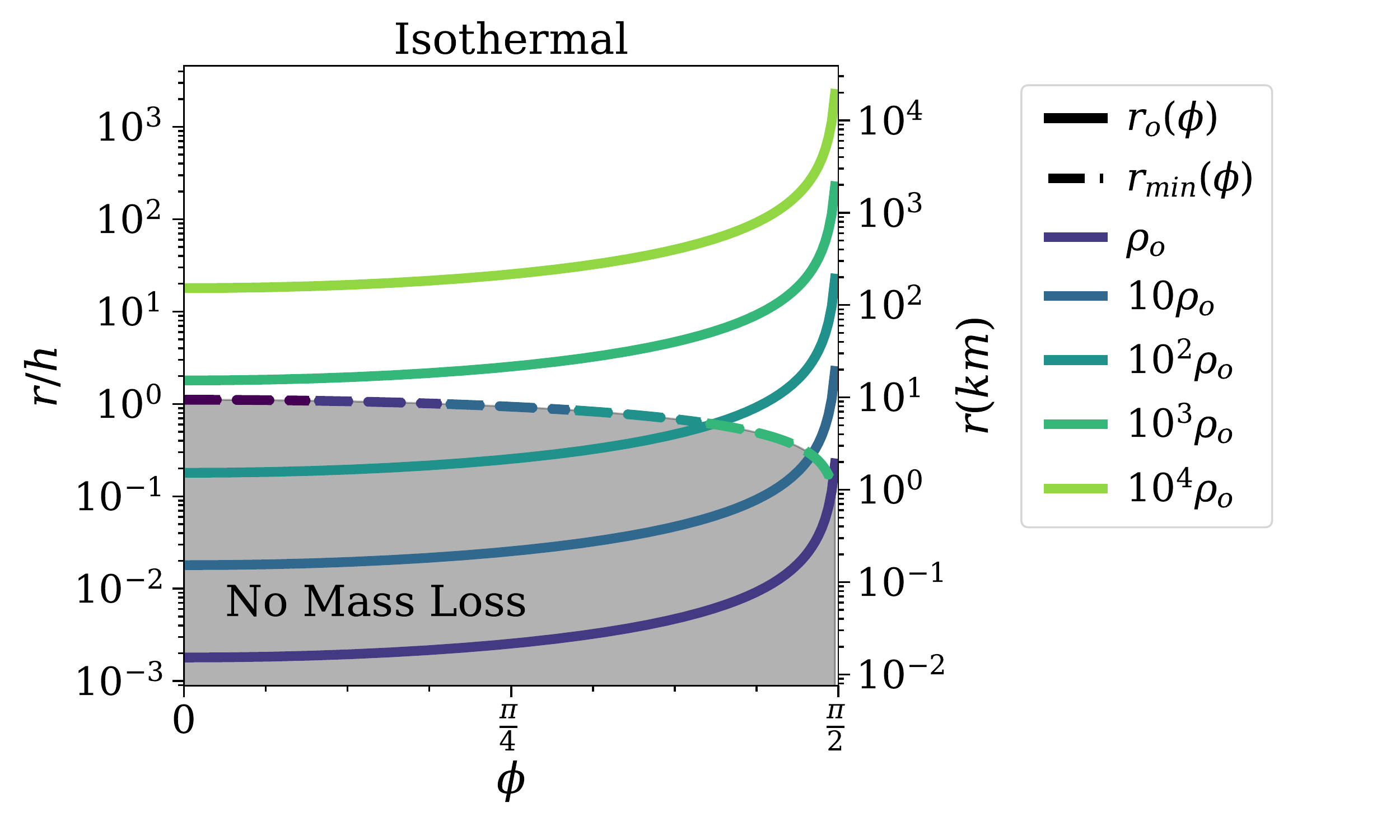}
    \includegraphics[width=0.49\textwidth]{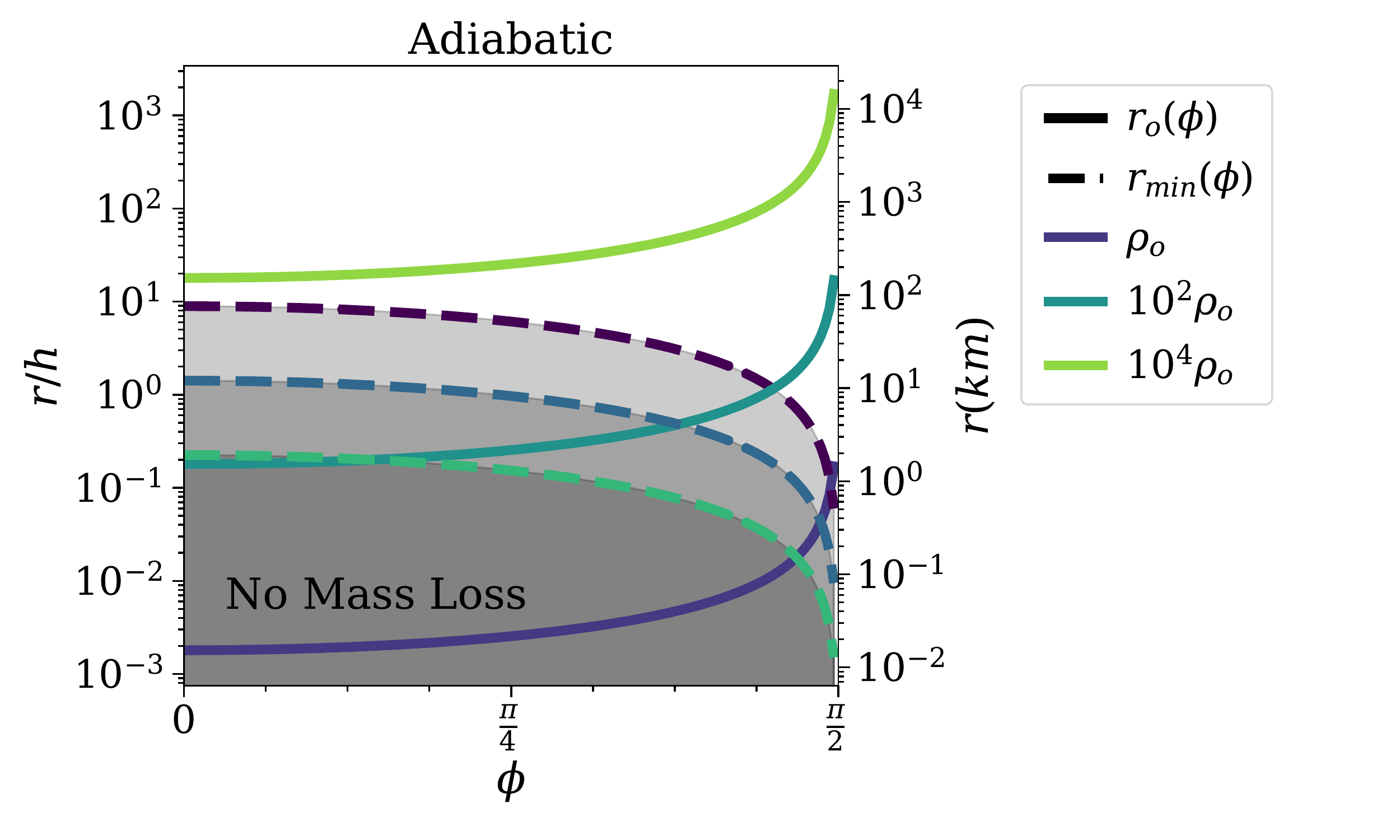}
   \caption{\textbf{Maximum impactor radius that will result in an aerial burst, $r_o$, as a function of trajectory angle $\phi$}, for various atmospheric densities ($\rho_o$). Here $\rho_o = 1.22$ kg/m$^3$, and we assume an impactor density of $\rho_i = 3300$ kg/m$^3$ and an initial velocity of $v_\infty/v_e = 2$. The top figure demonstrates the change in characteristic radii for an isothermal profile and the bottom shows the trends for an adiabatic profile. The dashed curves show $r_{min}(\phi)$, the minimum radius needed to eject any mass from a given trajectory angle. Note that $r_{min}$ is independent of atmospheric density in the isothermal case, and varies inversely with density in the adiabatic case. The solid curves show $r_o(\phi)$, the maximum impactor radius that will result in an aerial burst as opposed to a ground explosion. The $r_{min}$ and $r_o$ curves are colored by their corresponding atmospheric density. 
   The regime of mass loss by aerial bursts for a given atmospheric density is found above $r_{min}$ and below $r_o(\rho_o)$. Grey shading shows regions where no mass loss is possible (below $r_{min}$). Denser atmospheres increase $r_o$ at all trajectory angles, increasing the extent of the aerial burst mass loss regime in both impactor radius and trajectory angle. Because $r_{min}$ in the adiabatic case decreases with increasing atmospheric density, mass loss through aerial bursts can occur at relatively lower densities in the adiabatic atmosphere than in an isothermal atmosphere. }\label{rgroundrhos}
\end{figure}

In figure \ref{rgroundrhos}, we illustrate the range of planetesimal radii that can cause atmospheric-loss by aerial bursts for different atmospheric densities by plotting $r_o(\phi)$ and $r_{min}(\phi)$. In this example, we begin with an earth-like atmosphere, with $\rho_o=$ 1.22 kg/m$^3$ and $h=8$ km, and show the change in $r_o(\phi)$ as the atmospheric density is increased by consecutive factors of ten. We assume impactor densities of $\rho_i = 3300$ kg/m$^3$ and initial velocities $v_\infty=2v_e$. 

As calculated in section \ref{halfv}, the approximated form of the $r_o(\phi)$ curve is equivalent for both the adiabatic and isothermal profiles. $r_o(\phi) \propto \rho_o$, so the upper radius limit of aerial bursts $r_o$ increases with density as bodies must be larger in order to reach the ground before decelerating. $r_{min}(\phi)$ is independent of atmospheric density in the isothermal case, so the minimum radius limit of aerial bursts stays constant with increasing atmospheric density. As the atmospheric density increases, $r_o$ increases, and a larger range of aerial burst radii are able to eject atmospheric mass. 

Figure \ref{rgroundrhos} shows that for atmospheres with densities close to Earth's current atmospheric density, $r_o < r_{min}$ at all $\phi$ angles. In other words, only projectiles that reach the planet surface without decelerating will typically provide enough energy to result in atmospheric mass loss. As a result, for the current earth's atmosphere we find no significant atmospheric loss by aerial bursts. 

However, when the atmospheric density is increased by a factor of ten, $r_o > r_{min}$ for trajectories close to the horizontal ($\phi \sim \pi/2$). At this point, some aerial bursts can eject mass if they approach the planet's surface on a nearly horizontal path. Increasing the surface density by a factor of $10^4$ allows a range of aerial bursts approaching the planet on any path to eject mass. 

While the top panel of figure \ref{rgroundrhos} shows the $r_{min}$ curve for an isothermal atmosphere, we note that at high atmospheric densities we expect most atmospheres to be closer to adiabatic than isothermal. The bottom panel of figure \ref{rgroundrhos} demonstrates the changes in $r_{min}$ with atmospheric density for an adiabatic profile. As found in section \ref{characteristic radii}, the $r_{min}$ curve in the adiabatic case depends inversely on density ($r_{min, Adi} \propto \rho_o^{-2/5}$), so it would begin to intersect the $r_o$ curve at lower atmospheric densities than the isothermal $r_{min}$. This allows for mass loss by aerial bursts at relatively lower atmospheric densities than we found in the isothermal case. 

With the knowledge that aerial burst-induced mass loss occurs when $r_o (\phi)\leq r_{min} (\phi)$, we can calculate the minimum surface-level atmospheric density ($\rho_o$) required to eject mass through aerial bursts at a given trajectory angle ($\phi$). Equating the approximated forms of $r_o$ and $r_{min}$ and solving for $\rho_o$ results in

\begin{equation}\label{limit_phi}
\begin{split}
    \frac{\rho_{o}(\phi)_{min, Iso}}{\rho_i} &= \frac{64}{3} \left(\frac{\ln2}{\alpha C}\right)^{3/2} \frac{v_e}{v_\infty} \cos(\phi)^{3/2}\\
    \frac{\rho_{o}(\phi)_{min, Adi}}{\rho_i} &= \frac{32 \sqrt{2}}{3} \left(\frac{\ln2}{\alpha C}\right)^{3/2} \frac{v_e}{v_\infty} \cos(\phi)^{3/2},
\end{split}
\end{equation}
where $\rho_{o}(\phi)_{min}$ is the minimum atmospheric density at the planet surface required for mass loss by aerial bursts. Note that the limiting densities have no dependence on the scale height or planet radius in the $h \ll R, \phi \ll \pi/2$ approximation. A weak dependence of the limiting density on scale height and planet radius can be recovered if we evaluate the integral forms of $r_o$ and $r_{min}$ at $\phi = \pi/2$.

Taking the vertical limit as the most conservative indicator of whether aerial burst-induced mass loss is possible, we conclude that consideration of aerial bursts is necessary for atmospheres with densities of $\rho_{o}/\rho_i \gtrsim 0.4 v_e/v_{\infty}$.

\begin{figure}
    \centering
    \includegraphics[width=0.49\textwidth]{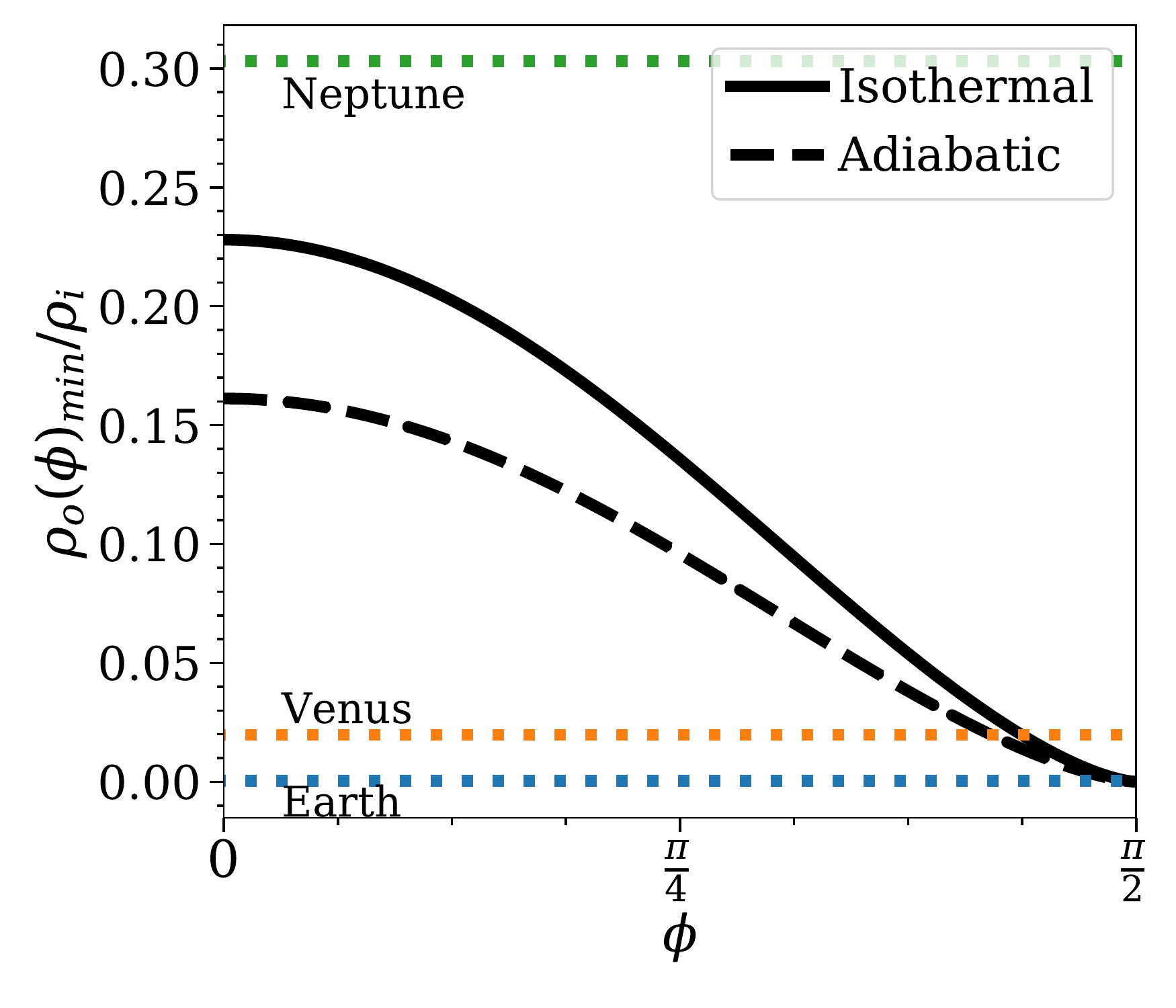}
   \caption{\textbf{Minimum atmospheric density required for the atmosphere to experience mass loss through aerial bursts} as a function of trajectory angle, $\phi$ (equation \ref{limit_phi}). Note, the atmospheric density given on the y-axis is normalized by the impactor density, $\rho_i$. The solid-black curve represents the isothermal profile, while the dashed-back curve corresponds to the adiabatic profile. Higher atmospheric densities are required for mass to be ejected by aerial bursts in isothermal atmospheres and by projectiles travelling vertically. As reference values, the atmospheric densities for Earth (blue), Venus (yellow) and Neptune (green) are shown as dotted lines.}\label{rhovphi}
\end{figure}

Figure \ref{rhovphi} illustrates the minimum densities in equation \ref{limit_phi} for isothermal and adiabatic atmospheres, assuming $\alpha = 9$, $C=1$, and $v_\infty= 2v_e$. We also show the approximate $\rho_o/\rho_i$ values for Earth, Venus, and the Neptune/sub-Neptune atmosphere, assuming impactor densities of $\rho_i = 3300$ kg/m$^3$. As seen in the example in figure \ref{rgroundrhos}, only denser atmospheres such as that of Venus or Neptune will experience mass loss by aerial bursts for projectiles on more vertical trajectories. 

Of particular note are the minimum atmospheric densities at $\phi = \pi/4$, as these trajectories are expected to represent the average collisions angle. We find, to experience mass loss by aerial bursts for impact angles of $\sim \pi/4$, atmospheres must have densities that are at least $\sim 10\%$ of the impactor density. 

\section{Contribution of aerial burst towards total mass loss}{\label{ground}}
At a given trajectory angle, all planetesimals with radii larger than $r_o(\phi)$ (equation \ref{rground}) will not decelerate significantly before reaching the planet surface, and will therefore result in ground explosions. We use the same prescription outlined in section \ref{methods} to describe ground explosions by simply setting the explosion height ($H$) in all equations to zero. Because ground explosion impactors do not decelerate significantly in the atmosphere, they will have impact velocities close to their initial velocities ($v_f \sim v_\infty$). 

The effects of ground explosions are described in detail in \cite{schlichting2015}; in this section we explore how aerial burst-induced mass loss compares to mass loss by ground explosions, and describe the overall differences between the mass loss curves calculated for isothermal versus adiabatic atmospheric density profiles. 


\subsection{Isothermal versus Adiabatic Atmospheres}
\begin{figure}
    \centering
    \includegraphics[width=0.49\textwidth]{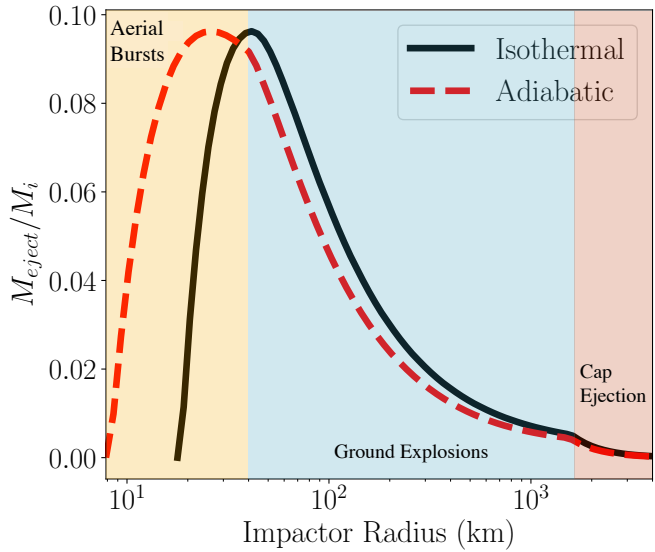}
   \caption{\textbf{Atmospheric mass loss per impactor mass for a sub-Neptune/Neptune like atmosphere} ($h = 19$ km, $\rho_o = 1000$ kg/m$^3$). We assume impactors of density $\rho_i = 3300$ kg/m$^3$ approaching the planet with trajectory angles of $\phi = \pi/4$. The solid-black curve shows mass loss assuming the atmosphere is isothermal; the dashed-red curve assumes an adiabatic atmospheric profile. Mass loss is separated into three distinct regimes based on the projectile radius: aerial bursts (shaded yellow region), ground explosions (shaded blue region), and ground explosions large enough to eject the full cap of atmosphere above the tangent plane (shaded red region). For sub-Neptune and Neptune-like atmospheres, aerial bursts can contribute significantly to atmospheric mass loss, especially in the adiabatic case and when accounting for the fact that smaller impactors are expected to be far more numerous than larger ones. }\label{neptunemami}
\end{figure}

In figure \ref{neptunemami} we show the application of both the aerial burst and ground explosion mass loss models to the same sub-Neptunian atmosphere discussed in section \ref{neptune} ($\rho_o = 1000$ kg/m$^3$, $h= 19$ km), assuming projectiles of density $3300$ kg/m$^3$ with trajectory angles of $\phi= \pi/4$. The curve spans three distinct categories of mass loss: aerial burst-induced mass loss at small impactor radii ($r_{min}<r<r_o$) shaded in yellow, ground explosions at intermediate radii ($r > r_o$) shaded in blue, and a tail at large radii where impactors reach the surface and the impacts are massive enough to eject the full cap of atmosphere above the tangent plane (where $\theta_m = \pi/2$) shaded in red. Note that at radii exceeding $\sim 1000$ km we may be entering the regime of giant impacts. 

As demonstrated in section \ref{characteristic radii}, figure \ref{neptunemami} shows that aerial burst-induced mass loss can occur at relatively smaller radii in an adiabatic atmosphere than in an equal mass isothermal atmosphere. This is largely due to the difference in half velocity heights between the two profiles (see figure \ref{halfheight}), which result in small particles decelerating relatively closer to the ground in adiabatic atmospheres. Because of the dependence on explosion height, the difference between isothermal and adiabatic mass loss is strongest in the aerial burst regime, and we see that projectiles can eject significantly more mass per impactor mass in the adiabatic case than bodies of the same radius in an isothermal atmosphere. Once impactors are large enough to reach the planet surface (the ground explosion regime), the gap between the mass loss curves is significantly reduced, and the two curves converge as we reach the cap ejection regime.

\subsection{Angle-Averaged Mass Loss}
\begin{figure}
    \centering
    \includegraphics[width=0.49\textwidth]{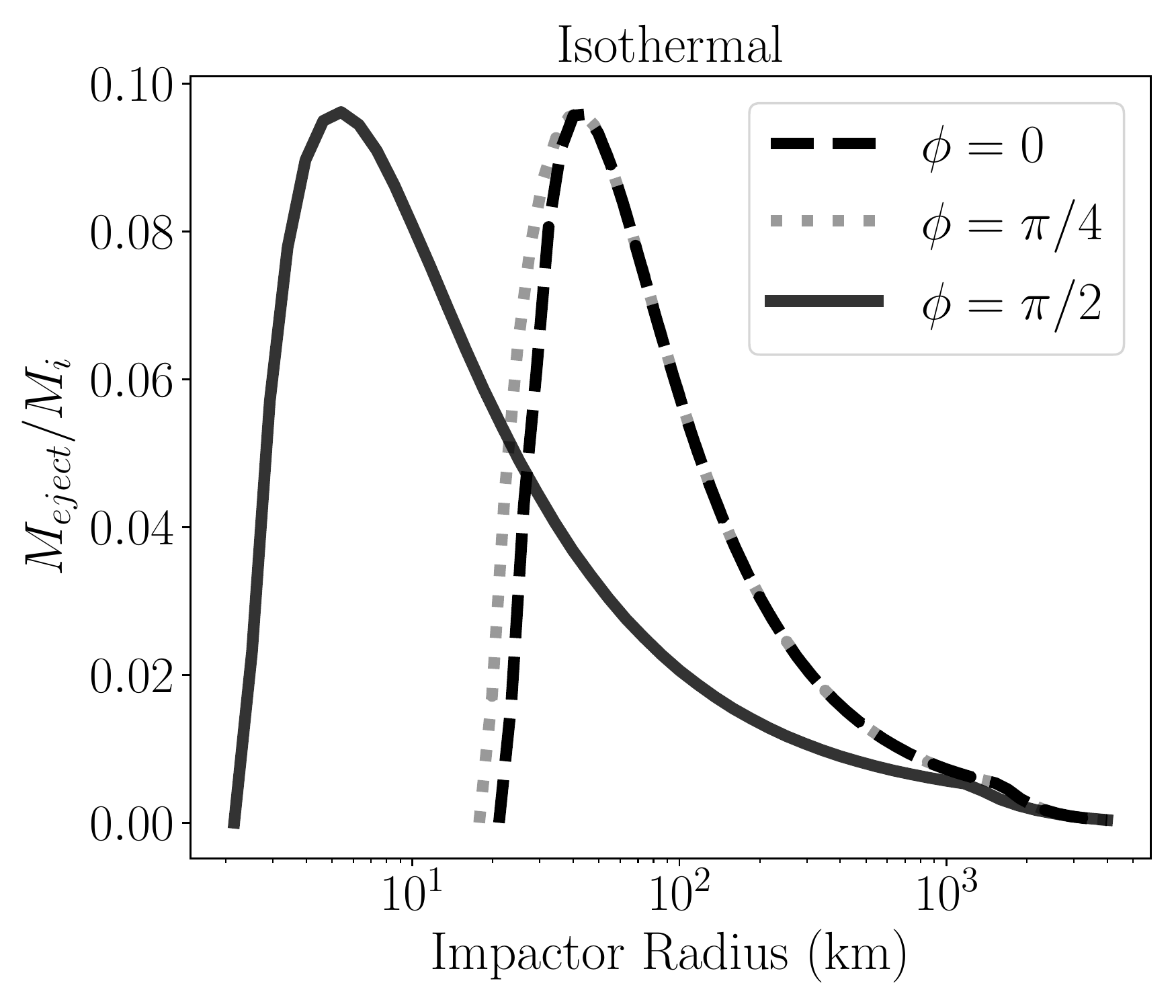}
    \includegraphics[width=0.49\textwidth]{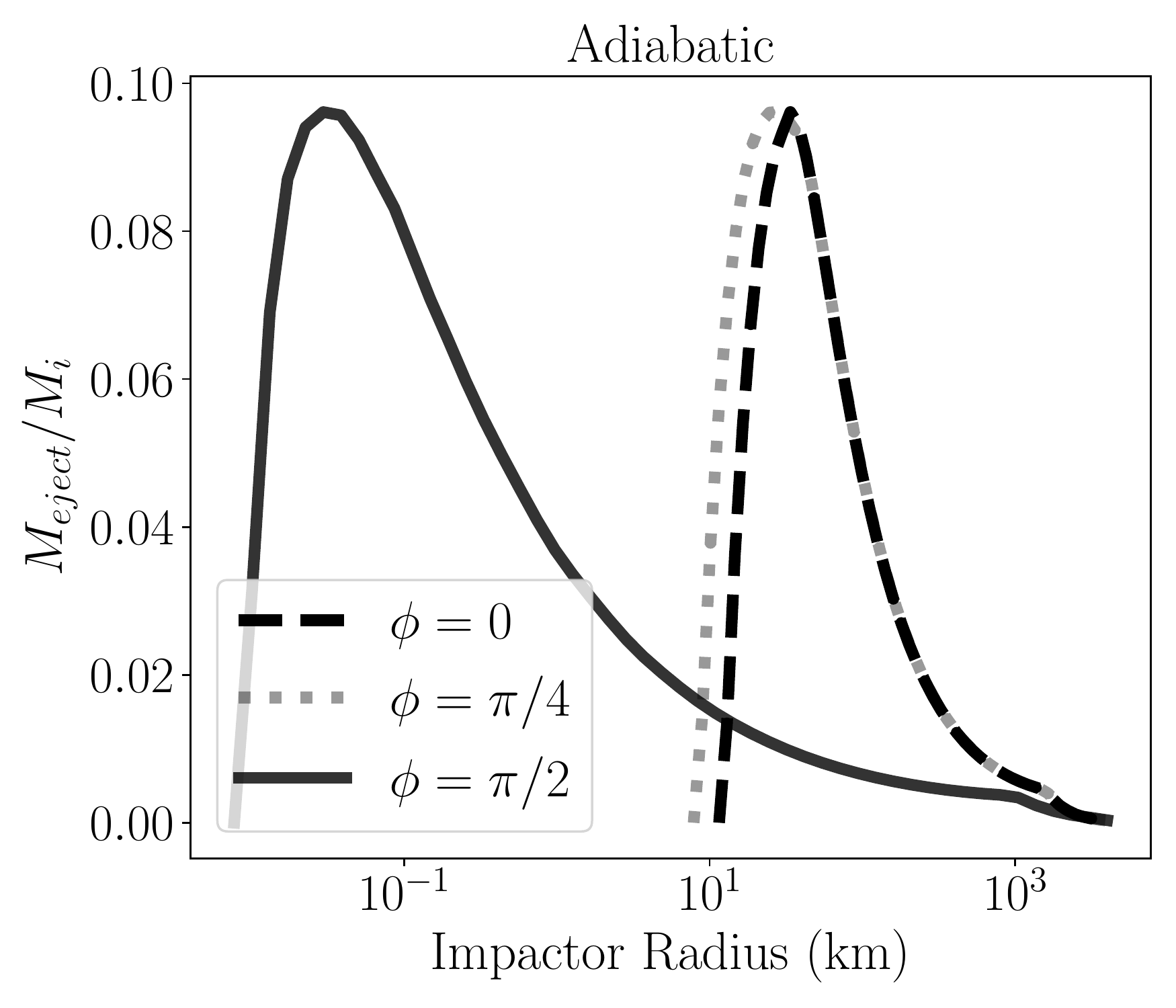}
   \caption{\textbf{Dependence of ejected mass per impactor mass on trajectory angle, $\phi$}, for $\phi=0$ (dashed line), $\phi=\pi/4$ (dotted line), and $\phi= \pi/2$ (solid line). We assume an atmosphere profile with $h = 19$ km and $\rho_o = 1000$ kg/m$^3$, and impactor densities of $\rho_i = 3300$ kg/m$^3$. Impactors travelling horizontally pass through more atmosphere than those vertically, so they decelerate higher in the atmosphere and result in a larger range of possible impactor radii that can lead to aerial bursts. 
   }\label{mami_angles}
\end{figure}

Figure \ref{mami_angles} shows how the atmospheric mass-loss in figure \ref{neptunemami} changes when the trajectory angle is varied from $\phi=\pi/4$, to $\phi =0$  and $\phi = \pi/2$. The general shape, progression from aerial burst to ground explosion to cap ejection, and relative difference between the isothermal and adiabatic results remain the same as in the $\phi=\pi/4$ case, with the curve simply extending towards smaller sizes as the trajectory angle increases. As described in section \ref{methods}, this is because projectiles moving horizontally along the tangent plane pass through relatively more atmosphere than those moving vertically, and will therefore decelerate more quickly, broadening the range of impactor radii that result in aerial bursts. This effect can also be seen in section \ref{characteristic radii}, which shows that $r_o$, the maximum radius of projectile that results in an aerial bursts, increases with trajectory angle $\phi$ while $r_{min}$ decreases with $\phi$.

\begin{figure}
    \centering
    \includegraphics[width=0.49\textwidth]{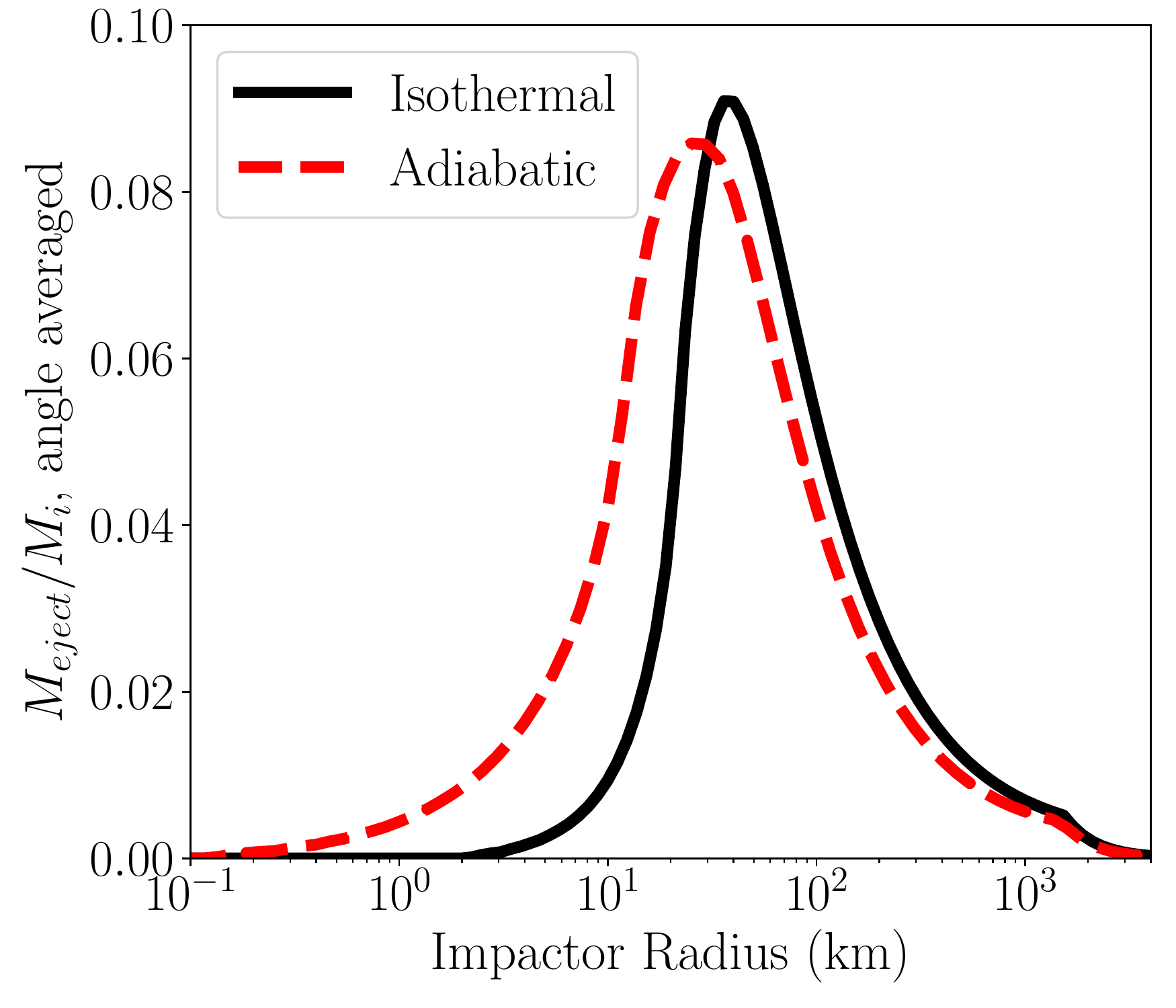}
   \caption{\textbf{Ejected mass per impactor mass averaged over all trajectory angles} ($\phi=0$ to $\pi/2$), for the isothermal and adiabatic profiles. We assume an atmospheric profile with $h = 19$ km and $\rho_o = 1000$ kg/m$^3$, and impactors of density $\rho_i = 3300$ kg/m$^3$. The peak in each curve is located at approximately the aerial burst/ground explosion transition, so that mass loss to the left of the peak is largely due to aerial bursts while the right corresponds mainly to ground explosions. The low efficiency tails of both curves at small impactor radii are largely due to the increased range of radii that can eject mass along horizontal trajectories. Meanwhile, the higher efficiency portion of the curve is dominated by the impactor radii corresponding to non-horizontal trajectories. }\label{mami_average}
\end{figure}

In figure \ref{mami_average}, we show the mass loss expected when we average over all trajectory angles, from $\phi=0$ to $\phi=\pi/2$, for a sub-Neptune/Neptune-like atmosphere. While the characteristic radii vary with trajectory angle, the peaks in mass loss efficiency for the adiabatic and isothermal curves are both located approximately at the aerial burst/ground explosion transition, similar to figure \ref{neptunemami}. Note that $r_{min}$ only decreases significantly at nearly-horizontal trajectory angles (figure \ref{mami_angles}), so that the range of radii that allow for mass loss is relatively similar at all other angles. As such, in the angle-averaged calculation the distribution of radii that result in the highest overall mass loss efficiencies is relatively narrow, as it is largely dominated by the radii that eject mass in the non-horizontal, $\phi \ll \pi/2$ cases. We therefore see a peak in mass loss efficiency at these radii, accompanied by a tail of low efficiencies corresponding to aerial bursts travelling the closer to the horizontal.

Overall, we find that aerial bursts are able to eject atmosphere with similar maximum efficiencies to ground explosions. These maximum efficiencies are a few to ten percent of atmospheric mass per impactor mass. Because of their smaller size, in a given impact, an aerial bursts will typically eject less mass than a single ground explosion. However, since smaller bodies are expected to be more numerous, their collective contribution to the overall atmospheric loss likely exceeds that by ground explosions.
 Assuming an impactor size distribution given by
 $dN \propto r^{-3.5} dr$ \citep{dohnanyi1969}, we find that the more frequent aerial bursts would result in a total atmospheric mass-loss that is larger by a factor of $(r_{ground}/r_{aerial})^{0.5}$ compared to ground explosion of similar efficiency. 

Therefore, given the cumulative strength of aerial bursts, they should be included in calculations of atmospheric mass-loss for thick atmospheres as defined in section \ref{constraints}.

\section{Discussion and Conclusions}{\label{conclusions}}


In this work we presented a simple analytical model describing aerial bursts and their associated atmospheric loss. Using this model, we have demonstrated that aerial bursts can lead to significant atmospheric loss for planets with sufficiently dense atmospheres, such as those expected for sub-Neptunes and Neptune-like planets. For a typical sub-Neptunian atmosphere, we find mass loss efficiencies of up to $\sim 10 \%$ of the impactor mass. Specifically, the main conclusions of this paper are as follows:
\begin{enumerate}
    \item Mass loss from aerial bursts becomes significant when the maximum impactor size that will lead to an aerial burst rather than a ground explosion, $r_o$, is larger than the minimum impactor size needed to achieve atmospheric loss, $r_{min}$. For vertical trajectories, which give the most stringent limit, this condition is approximately satisfied when $\rho_o/\rho_i \gtrsim 0.4 v_e/v_\infty$, which implies atmospheric densities need to be comparable to impactor densities for impactor velocities that are a few times the escape velocity of the planet. 
    As a result, aerial bursts are not expected to significantly contribute to the atmospheric mass-loss history of Earth and Earth-like planets, but are expected to play an important role for planets like Neptune with significant atmospheres.
    \item The range of impactor radii resulting in aerial burst-induced mass loss, $r_o-r_{min}$, depends on the ratio of the density of the atmosphere to the impactor density, and varies with the trajectory angle, such that the range increases for larger impact angels and larger atmospheric densities.
    \item Because adiabatic atmospheres have larger densities for a given height above the ground within in the first scale height compared to isothermal ones, the range of impactor radii that result in aerial burst-induced mass loss is larger in adiabatic atmospheres than isothermal atmospheres of equivalent total mass, scale height, and atmospheric surface density. Similarly, aerial burst impactors of equal size will eject relatively more mass in an adiabatic atmosphere than in the isothermal atmosphere of equivalent total mass and scale height.
    \item Increasing the density or initial velocity of impactors will increase the expected mass loss per impactor mass in both the ground explosions and in aerial bursts.
    \item The difference between isothermal and adiabatic atmospheric density profiles matters most in the aerial burst-induced mass loss regime. The mass lost per impactor mass is similar for both profiles in the ground explosion regime. 
    \item For Neptune-like atmospheres, the atmospheric mass ejected per impactor mass by aerial bursts is comparable to that lost by ground explosions. Therefore, for impactors following a Dohnanyi size distribution, overall loss by aerial busts is expected to exceed that by ground explosions by a factor of $(r_{ground}/r_{aerial})^{0.5}$.
\end{enumerate}

In addition to the contributions towards overall atmosphere mass loss discussed thus far, the knowledge that a large number of projectiles will result in aerial bursts as opposed to ground explosions may be important in determining the final compositions of atmospheres. 

Because aerial bursts are significantly decelerated high in the atmosphere, they may be useful for depositing volatiles on a planet's surface. In the sub-Neptune atmosphere ($\rho_o \sim 1000$ kg/m$^3$), we find that impactors with radii up to tens of kilometers could be decelerated before they reach the ground. These objects could therefore reach the surface of the planet without creating a large crater and destroying the majority of the impactor. Hydrodynamical simulations also show that for impactors with the density of asteroids ($\sim 3300$ kg/m$^3$), little to none of the impactor's own mass will be ejected over the course of its path through the atmosphere (\citealt{shuvalov2014}). As such, these aerial bursts could trade small amounts of atmospheric loss for delivery of their own volatiles to the planet. 

\section*{Acknowledgements}
HES thanks Mark Wyatt for suggesting to work on this topic. This research was supported by the National Aeronautics and Space Administration under grant No. 80NSSC18K0828. 

\section*{Data Availability}
No new data were generated or analysed in support of this research.



\bibliographystyle{mnras}
\bibliography{bibli} 



\bsp	
\label{lastpage}
\end{document}